\documentclass[10pt, format=manuscript,nonacm]{acmart}
\usepackage{graphicx}
\usepackage{makecell}
\usepackage{array}
\usepackage{pdflscape}
\usepackage{booktabs, makecell, multirow, tabularx,
	threeparttable, tabulary}
\usepackage{array}
\usepackage{subfigure}
\usepackage{fancyhdr}
\AtBeginDocument{%
  \providecommand\BibTeX{{%
    \normalfont B\kern-0.5em{\scshape i\kern-0.25em b}\kern-0.8em\TeX}}}

\setcopyright{acmcopyright}
\copyrightyear{2022}
\acmYear{2022}
\acmDOI{10.1145/1122445.1122456}

\acmJournal{JACM}
\acmVolume{37}
\acmNumber{4}
\acmArticle{111}
\acmMonth{8}

\begin{document}

\title{From “study with me” to study with you: how activities of Study With Me livestream on Bilibili facilitate SRL community}


%
\author{Ge Wang}
\affiliation{%
	\institution{University of Science and Technology of China}
	\streetaddress{University of Science and Technology of China,
		No.96, JinZhai Road Baohe District}
	\city{Hefei}
	\country{China}}
\email{oga@mail.ustc.edu.cn}

\author{Yanxiang Zhang}
\affiliation{%
	\institution{University of Science and Technology of China}
	\streetaddress{University of Science and Technology of China,
		No.96, JinZhai Road Baohe District}
	\city{Hefei}
	\country{China}}
\email{petrel@ustc.edu.cn}

\renewcommand{\shortauthors}{Ge Wang and Yanxiang Zhang}

\begin{abstract}
  It has become a trend to use study with me (SWM) Livestream to create a personalized study ambiance. However, we still have little understanding of the activities of SWM livestream and the streamer’s motivation to produce SWM livestream. This paper provides an overview of the activities and how streamers regulate these activities of SWM livestream on a Chinese popular User Generated Content(UGC) website, Bilibili. We observed the number and popularity of the SWM livestreams and analyzed 800 livestreams to understand the streamers' study goals. We analyzed 20 SWM livestreams in detail and interviewed 12 streamers and 10 viewers to understand the activities and the streamer’s motivation. Based on the prior research on computer-supported collaborative self-regulated learning (SRL), we identified that SRL community needs cognitive supporting activities and social supporting activities. We used the theoretical framework of SRL and sense of community (SOC) to understand how SWM livestream activities facilitated these two dimensions. We found that streamers produced SWM livestream to seek supervision, find like-minded stud partners and help and company others. Streamers don’t interact or instruct with the viewers directly but use chat-bot and autonomous interaction to alleviated the interaction burden. Unique sessions like checking-in and study progress reporting promote the viewers’ social presence, promoting SOC, and enhancing their engagement. Strict rules and punishment are widely used to concentrate the members on study and contribute to positive atmosphere. We also found that SWM livestream often disappears when the examination is done and the streamer faces doubts on motivation and appearance. These findings suggest that SRL community can provide cognitive and socioemotional support for lonely learners to stick to a long-term study. The activities and streamer’s practice inspired how streamers can focus on contemplative efforts while controlling the interaction.
\end{abstract}



\maketitle
\thispagestyle{fancy} 
\fancyfoot[LE,RO]{Manuscript submitted to ACM}
\section{Introduction}
Books on the desk, writing or typing hands, an electronic timer, a highly focused streamer, and no direct interaction nor instruction, these are the daily live session a “study with me” (SWM) streamer would show on the screen. In 2018, a South Korean streamer nicknamed the Bot - No - Jam (t) has attracted more than 321000 subscribers on his YouTube channel. He posted SWM livestream on YouTube, and he called this "Study with Me". The duration of this livestream is very long. Each learning session or livestream lasts an average of six hours \cite{3}. On YouTube, SWM livestream are gaining increasing attention \cite{youtube1}. Another similar livestream on YouTube named "Lofi hip hop radio" is a looping animation with chilling music showing a girl studying at home \cite{2}. Surprisingly, in all kinds of reports, streamers and viewers of SWM livestream said it improved their study motivation and provided them with supervision, a sense of companionship, and competition \cite{1,3}.

In China, the YouTube-like user generated content (UGC) video website Bibibili is an essential base for SWM livestream. cctv.com (China Central Television) released a news piece, \textit{Do you know that this generation of young people would love to study on Bilibili} on April 17, 2019 \cite{1}. The news described the current situation of netizens studying on Bilibili. Data from Bilibili showed that 18.27 million people had studied on Bilibili in 2018. The kind of livestream hash-tagged \#Study with Me\# has become the category with the longest livestream time of Bilibili. In 2018, the total duration of SWM livestream reached 1.46 million hours, and the number of them reached 1.03 million \cite{1}. On Bilibili, there can be a maximum of 2,000 to 3,000 SWM livestream per day. Popular SWM streamers can have hundreds of thousands of subscribers. 
\begin{figure}[h]
	\centering
	\subfigure[A SWM livestream on Bilibili.]{
		\begin{minipage}{15cm}
			\centering
			\includegraphics[width=0.6\linewidth]{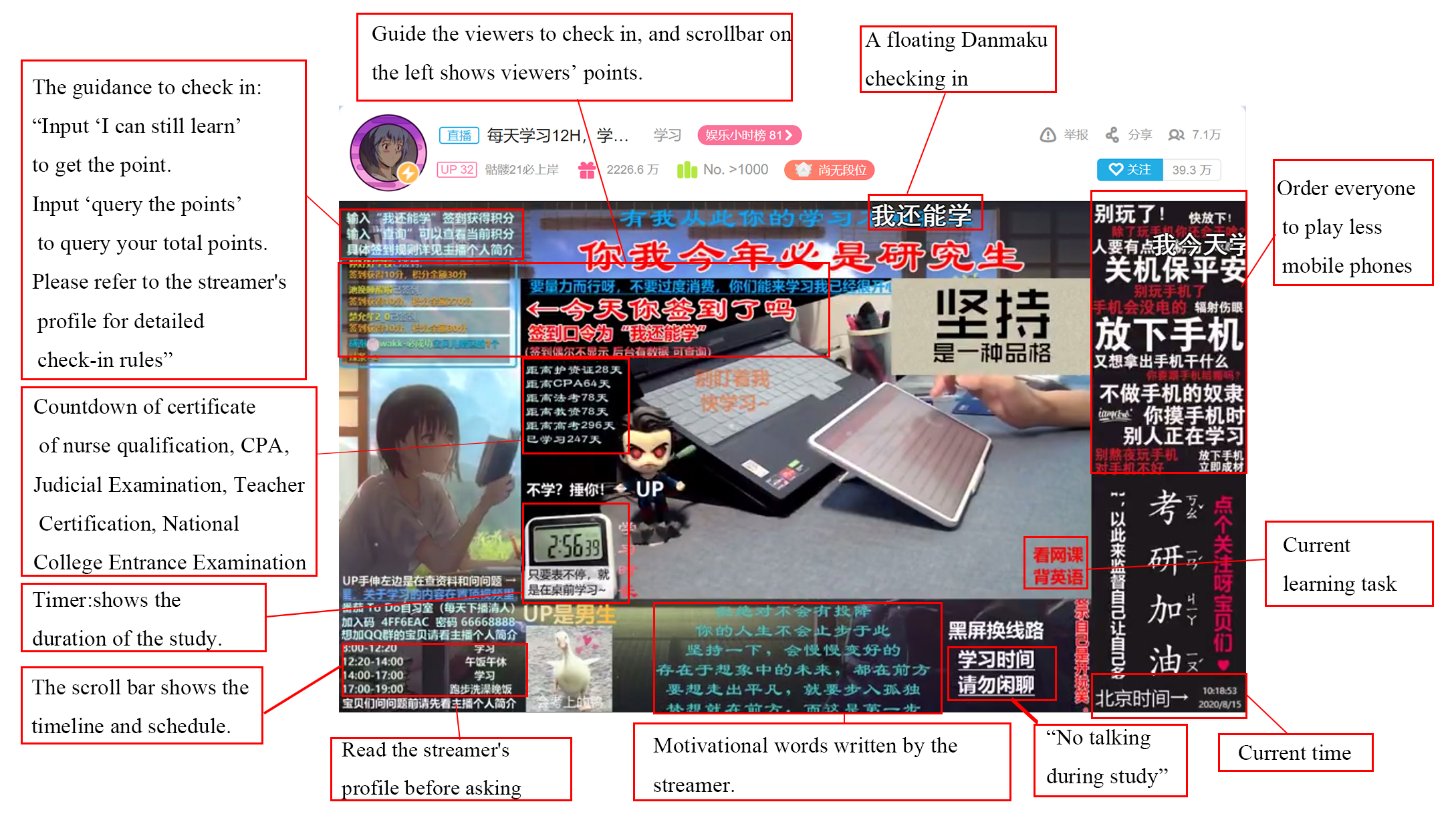}
		\end{minipage}
	}

	\subfigure[A typical SWM livestream on Youtube.]{
		\begin{minipage}{15cm}
			\centering
			\includegraphics[width=0.6\linewidth]{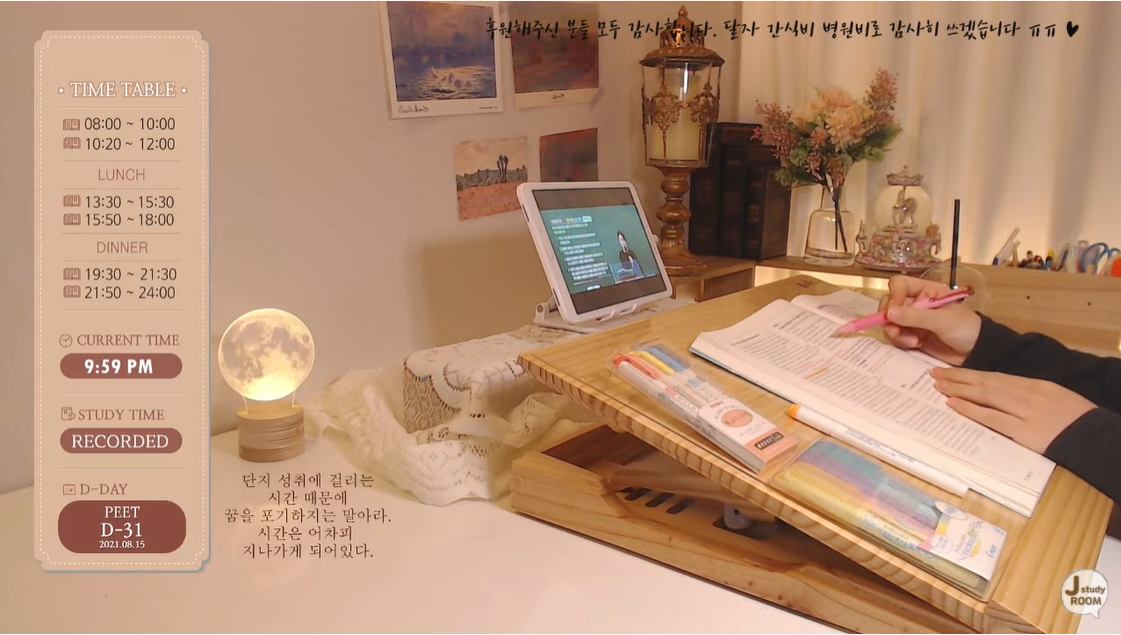}
		\end{minipage}
	}
	\caption{The snapshot of a SWM livestream on Bilibili (the layout of this livestream contain almost all the typical activities) and YouTube.}
	\label{snapshot}
\end{figure}

Despite the contribution of a personalized study ambiance, we still have little understanding of the practices and activities of SWM livestream and the influence of these activities on SRL community. In prior research on SWM videos, they described SWM videos as a kind of video that shows the recorded real learning sessions of the uploaders. The uploaders don’t interact nor instruct directly with the viewers in the video. They found SWM video is a new way to study in the presence of others and that watching SWM videos is an environmental regulation of self-regulated learning (SRL) \cite{lee2021personalizing}. But the potential of SWM livestream for a SRL community is still underexplored. The research about knowledge-sharing-focused livestreams, such as creative livestream \cite{16}, programming livestream\cite{15}, and intangible cultural heritage livestream \cite{34}, has shown how these livestream practices help the streamer and viewers promote culture, mentor each other, get inspired, and develop computer supported collaborative learning (CSCL). However, SWM livestream shows different characteristics from knowledge-sharing livestream or CSCL community because it doesn’t focus on pedagogical process but support self-regulated learning (SRL). In China, with the support of local social network and other SRL-support software, SWM livestream in local environment presents various unique activities as shown in Fig.\ref{snapshot}, and can facilitate an SRL community. Therefore, we regard SWM livestream as a computer supported collaborative SRL (CSCSRL) and it is underexplored how the practices of SWM livestream support such a community and facilitate SRL.

In this paper, we explored how SWM livestream support SRL community. To understand the activities and practices in SWM livestream, we conducted an observation of SWM livestream on Bilibili for a year and a half. In combination with titles and snapshots, we analyzed the contents of 800 SWM livestreams to understand the study goals of streamers. And we conducted a more in-depth content analysis on 20 SWM livestream to understand the organization, management and regulation of the activities of SWM livestream.

To have a deep understanding of the activities of SWM livestream, we conducted interviews with 12 streamers and 10 viewers. Combined with observation and interview, we found that 1) streamers produced SWM livestream to seek supervision, find like-minded study partners, and help with other’s SRL; 2) sharing the study experience and plan helped members to reflect on their study and regulated their study plan and behavior; 3) checking in and study progress report promote the viewers’ social presence, which can enhance their engagement, help with SRL and facilitate SOC. 4) strict activity rules, limitation of the topics, and punishment mechanism concentrated the members on study and contribute to community security; 5) chat-bot and autonomous interaction of the viewers are widely used in SWM livestream, which alleviated the streamers interaction burden but limit the activity of interaction. 6) streamers expand the interaction beyond Bilibili and set up private fans group and multi-person virtual study room to facilitate SRL.

The contributions of this work are, thus, an observation and interview-based study that identified \romannumeral1) the practices and activities of SWM livestream and how this differs from other entertainment livestream or knowledge-sharing-focused livestreams; \romannumeral2) the motivations of SWM streamers, \romannumeral3)how streamers organize, manage and regulate the activities and facilitate SRL and SOC.

\section{Related work and background}
\begin{figure}[htb!]
	\centering
	\includegraphics[width=0.8\linewidth]{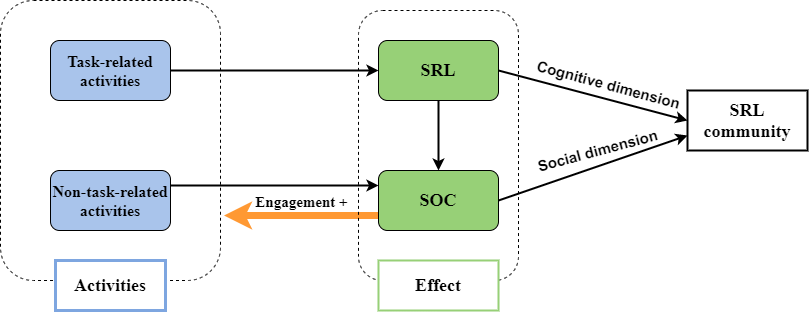}
	\caption{Theoretical framework that explains the relationship among the activities, SRL, SOC and SRL community.}
	\Description{framework}
	\label{framework}
\end{figure}

Research on computer-supported collaborative SRL, sense of community, and livestream provide the theoretical framework for our research. After summarizing and combing the literature on these three areas, we developed a framework for understanding how activities in livestream support SRL and SOC in the SRL community. The social activities in two dimensions, task-related (cognitive) and non-task-related (social) mainly contribute to SRL and SOC respectively. SRL contributes to SOC because it meet the members integration and fulfillment. And SOC promotes the members engagement. Finally, SRL and SOC facilitate a CSCSRL community for congnitive and social dimention respectively. This framework is shown in Fig.\ref{framework}.
\subsection{Computer-supported collaborative Self-regulated learning (CSCSRL)}
Self-regulated learning (SRL), refers to individual learners taking metacognitive control of cognitive, behavioral, motivational and emotional conditions/states through iterative processes of planning, monitoring, control, and change. The ability of SRL is becoming increasingly important 21st century because many students are learning from online resources alone but lack autonomous to manage their learning pace \cite{schunk2012motivation}. SRL is not easy and needs to be both learned and also often supported with self-regulation tools and/or environments \cite{hadwin2010innovative}. 

Computer supported collaborative SRL community needs to be satisfied in two dimensions, one is to support members’ SRL cognitively, and the other is to support a sound social space for successful collaboration. Prior research on computer supported collaborative SRL support was mainly in CSCL environment. These researches have two directions: one focus on technological and educational affordance and the other focus on social affordance \cite{jarvela2015enhancing}. The former is developing computer based tools to support self-regulated learning. Some tools support task notification\cite{forest,freedom,stayfocused}, time management\cite{cirillo2006pomodoro,pomofocus,rescuetime}, productivity-tracking \cite{kim2016timeaware,rooksby2016personal,whittaker2016don,rooksby2016personal}to improve self-awareness, goal-setting, and planning that depend on the learning phase \cite{bannert2012supporting}. These tools aim at activating learners’ SRL skills individually. In China, a trend of SRL-supporting tools is introducing community-support functions. Users can build virtual study rooms, where they can study together or track every one’s learning status by study hour ranking, multi-person livestream or virtual images \cite{timing,fqtodo,costudy}. The second line of research focus on group awareness and sociability support. They tend to increase social interaction in the cognitive dimension and socioemotional dimension to build a sound social space for SRL. The members in such an environment should feel relatedness, group cohesiveness, trust, and respect for each other \cite{jarvela2015enhancing}. 

Researchers also found that livestream facilitates study-related collaboration. From the social perspective, the streamers can use livestream to build a learning community or knowledge-sharing-based community, where the members can get mentorship, inspiration, companionship and sense of belonging \cite{chen2021towards,16,34,15}. From the technological perspective, the prior research on SWM video argued that watching SWM video (or livestream) is a kind of productivity tool and environment control of SRL, which helped the viewers focus on their studies and avoid distractions \cite{lee2021personalizing}.

The practice of SRL supporting community is still underexplored since the prior research mainly focus on SRL in CSCL, which emphasis on the result of co-construction of knowledge. Our work aims at having a nuanced understanding of SWM livestream, which is in non-pedagogical context and focus on study behavior regulation. From the perspective of technological and educational affordance, we will understand how SWM streamer’s and viewer’s practice support SRL. From the perspective of social affordance, we will explore how do activities in SWM livestream promote SOC of the members.
\subsection{Social space and Sense of Community (SOC)}
Researchers have realized that a sound social space is the foundation of collaboration \cite{settle2016improving}. Prior research emphasized that students need to trust each other, feel a sense of warmth and belonging, share a sense of community, have common goals, and feel close to each other before they will engage willfully in collaboration \cite{settle2016improving,kreijns2013social}. An important concept supports social space is the sense of community (SOC), which is a feeling that members have of belonging, a feeling that members matter to one another and to the group, and a shared faith that members’ needs will be met through their commitment to be together \cite{mcmillan1986sense}. McMillan’s definition of SOC has four elements: membership (sense of belonging), influence (being influential in the community and influenced by the community), integration and fulfillment of needs (people meet others’ needs while they meet their own) and shared emotional connection (based on a shared history) \cite{mcmillan1986sense}.

To develop a sound social space, social interaction in cognitive and socioemotional dimensions should be increased \cite{jarvela2015enhancing,sun2019distance}, the former refers to on-task, educational, instructional, or pedagogical interaction; the latter   refers to non-task, off-task, social, or non-pedagogical interaction\cite{abedin2012nontask,gilbert1998building}. Combined with the context of SRL, the cognitive dimension is task-related and SRL-supporting interaction. The socioemotional dimension is non-task-related and SOC-supporting interaction.

Much efforts have been done to explore how to enhance the interactions in the above two dimensions. Some research revealed the importance of the regulation of social activities in learning communities. Some explore how to enhance activities in the above two dimensions. For example, some research used Multimedia Tools to Enrich Interactions in Live Streaming language class to enrich learning experience and improve students’ performance \cite{8}, and some designed linked-courses learning community for development majors to reduce students’ social isolation and to improve retention rates and academic performance\cite{settle2016improving}.
Our work aims to understand how task-related and non-task related activities are organized and regulated in SWM livestream. And further explore the relationship between SWM livestream practice, SOC and SRL.
\subsection{Strategies to foster livestream community}
Livestream provides strong financial and social incentives for streamers to developing a follower base. Much research have revealed that streamers often want to foster their community \cite{34,16,15,35,23,44}

To attract viewers to watch and keep them engaged, the streamers have to curate the content and interaction during the livestream carefully. To attract the viewer’s attention, effective streamers must find topics that attract and retain a viewership based on shared interests, entertainment capacity, and the streamer’s technical skill \cite{15}. Even if it is a knowledge-sharing livestream, they should also pay attention to the balance between knowledge and entertainment \cite{15,16,34,chen2021towards}. In addition to the content, the streamer has to construct an attractive on-stream personality and establish their social presence. They often construct a welcoming personality by greeting viewers, responding messages in their stream chat in a timely manner \cite{15} and show their unique personality \cite{23,35}.
In terms of the interaction, the online peer work and collaboration can be synchronous (video meeting, chat, livestream etc.) and asynchronous (e-mail, discussion board, etc.). A research studied the synchronous and asynchronous communication in E-learning and found that synchronous communication calls for personal participation, which increase arousal, motivation and convergence on meaning. While asynchronous communication calls for cognitive participation, which increase reflection and ability to process information \cite{hrastinski2008asynchronous}.
Thus, streamers use various synchronous interaction to keep viewers engaged and entertained. They used voting (the viewers vote on what the streamer should do next), contest, request, Q\&A sessions, audience participation games, thanks the viewers gifts, battle session, etc \cite{lee2018tip,15}. 

However, in some knowledge-sharing-focused livestream, the tasks in livestream often need more cognitive work and streamers often have trouble focus their work while chatting with the viewers. So, they use moderators or host, install widgets, use chat bots and automatic respond to the message to alleviate their interaction pressure. Besides, the streamers also use some asynchronous interactions, for example, asking the viewer’s contribution to the project and gave the viewers a daily challenge \cite{15}\cite{16}. 
Our work will explore how SWM streamers manage the content and interaction in their livestream. According to the challenges faced by the current non-entertainment-focused livestream, this work will discuss the implication for the streamers and platforms.
\section{research object}
To support our motivation for this study and to help us understand the practices in SWM livestream, we introduced the background and characteristics of Bilibili, the difference between SWM livestream and video, the reason why we choose SWM livestream on Bilibili as the research object, as well as the functions that streamers and viewers can use in Bilibili.
\subsection{Bilibili}
Bilibili is a popular UGC content website in China. As a highly clustered cultural community and video platform for young generations in China, young people with different interests can find their circle at Bilibili, which has become a multi-cultural community covering more than 7,000 interest circles \cite{bilibili}. According to Bilibili's first-quarter financial report released on May 13, 2021, the average monthly active user (MAU) reached 223 million, and the mobile MAU reached 208 million. Bilibili averaged 60 million daily active users (DAUS) in the first quarter \cite{caibao}. In prior research, Bilibili was described \cite{jia2017analysis,lu2019user} as a UGC website similar to YouTube, but with more social functions. Like traditional UGC sites, BiliBili users can watch and share videos, vote and comment on videos, subscribe to channels (a series of videos), and share their updates. Additionally, Bilibili offers three (unique) social features :(\romannumeral1) non-interactive users follow, allowing users to follow each other for social purposes or just to get interested in video updates. (\romannumeral2) chat playback, known as "Danmaku" in Bilibili, is comments that fly across the screen, and users who watch the video post Danmaku comments. (\romannumeral3)Virtual currencies were donated by users to the uploaders to thank the uploaders for their contributions. At Bilibili, virtual currencies (named coin) are useful in various situations, including upgrading users' identities and exchanging new emojis.
\subsection{The difference between SWM livestream and SWM video}
The main body of the SWM livestream and SWM video is the same, the creator’s study footage without direct instruction with the viewers. However, there are some difference between these two subjects. Firstly, SWM livestream is in real time but the SWM video is prerecorded, which means that the streamers are studying with the viewers at the time. This provide opportunity for a community because the streamers are viewers are facing similar challenges in study and they can share their experience and really study together. Secondly, SWM livestream shows the whole process of streamers' study process at a normal speed but the video is often part of the study footage segment and sometimes accelerate the study process. Therefore, SWM livestream shows a more real study process and the image of streamers.
\subsection{Why we choose the SWM livestream on Bilibili?}
First of all, Bilibili has the largest number of SWM livestream in China and also has a considerable user base \cite{1}. As the news reported, students learning on Bilibili has become a typical phonomenon in China and the duration of SWM livestreaming on Bilibili was considerable. Due to the gradual expansion of SWM livestream, it also has its category on Bilibili. The SWM livestream on Bilibili is in the “study companion” subcategory under “study” category in livestream feature. There are about thousands of people streaming their study footage each day. Other mainstream live platforms in China, such as Douyu and Huya, are mainly for gaming livestreams and talent livestreams, and there are only fewer than ten people producing SWM livestream. Therefore, we take the SWM livestream on Bilibili as the research object.

Furthermore, Bilibili also provides more opportunities for the expansion of community because it is a local UGC platform. The streamers and viewers of SWM Livestream on Bilibili are all Chinese learners. The same culture and environment for examninations provide the soil for the formation of SRL communities. With the help of other local SNS such as QQ and WeChat, SWM Livestream is able to thrive and develop its own characteristics in the local Chinese environment. Therefore, the SWM Livestream on Bilibili provides rich and unique materials to understand the practice of SWM Livestream.
\subsection{What streamers can do on Bilibili?}
The users can produce livestream on Bilibili after authenticating their identity. Bilibili has it’s own software for livestream. When livestreaming the streamers can set the layout of the livestream and add words and picture on the livestream screen. They can write down an introduction of their livestream and other users can see it below the livestream. They can see the viewers’ Danmaku, mute the viewers (all the viewers can not send Danmaku), block the viewers (specific users can not send Danmakus) and block Danmakus that include specific words.
To increase self-disclosure, streamers often use some general functions for every user on Bilibili.

\textit{Moment (Dongtai)} is an ins-like function. Users can post picture with words (or only words) to show their lives and express their feelings.

\textit{Column} is a blog-like function. Users can use this function to post long articles. The column support to add pictures and in-site videos.

\textit{Private massage} is the chat function that the users can send message to each other. Only the sender and the receiver can see the message.

\subsection{What livestream viewers can do on Bilibili?}
When watching livestreams, the viewers can also send Danmaku and donate virtual gifts to the streamers. The viewers can donate free virtual gifts to the streamers and pay for the virtual currency called "golden melon seed" and "silver melon seed" to buy other gifts to the streamer.
They can also send private message to the streamer. Besides, there are some privileges for the viewers. The viewers can purchase the title such as "captain", "admiral", and "governor", then they can get the corresponding privilege such as an ID card showing their privilege identity,the privilege to send longer Danmakus and Danmakus on the bottom of the screen, particular color for their Danmakus, special chat bubble, etc. Part of the money they paid for the title goes to the streamers. 
\section{research methods}
To have an overview of the activities of SWM livestream and deeply understand the activities and streamer’s motivation, we conducted observation of the SWM livestreams on Bilibili and semi-structured interviews with 12 streamers and 10 viewers.
\subsection{Observation}
To get a general understanding of the streamers’ SWM livestream practice. We did an observation of the SWM livestream on Bilibili. Before the observation of specific streamers, we had a general overview of the topics of SWM livestream on Bilibili to understand who are streaming and what are they learning. We browsed all the livestreams in the “study companion” (Peibanxuexi) directory at first. To have an overview of the SWM streamers, we used python crawler 3 times a day for 7 days to obtain the number of SWM livestreams and number of currently-watching viewers of each livestream. We used crawler to collect the titles and snapshot of the top 800 livestreams in "study companion" directory.The streamers often disclose their study goals in the title or on the screen. Therefore, we did a content analysis of the titles and the snapshot of these livestreams to understand what the streamers are studying for.

After having a general understanding of the livestream. We randomly selected 20 streams form the above 800 livestreams for detailed analysis.In addition to the livestream video, we collected the introduction and the streamers’ homepage of these 20 streamers on Bilibili. Because we noticed that the introduction and homepage contain much information about the streamers themselves and how to participate in the stream and interact with the streamer. This information provides rich content for us to understand the streamers’ practice. One author observed these information to identify 
novel activities, and the other reviewed the result. 

\subsection{Interview research}
Observations were supplemented with semi-structured interviews with streamers and viewers to gain further insight into their practices. We conducted interviews with 12 streamers and 10 viewers of SWM livestream. We recruited our participants via purposive sampling combined with snowball sampling. The first 5 streamers were recruited from the streamers we have observed. The rest of the streamers and all of the viewers were recruited through snowball sampling. We reached out to interview candidates by Bilibili private message with a short description of our research purpose and a background survey about their live streaming experience and demographic information. Based on their responses, we selected our final interviewees (refers to both streamers and viewers) to ensure the relevance of their experiences, as well as consider diversity in their study goals.

We designed the interview based on our research question and the observation. In the interview of streamers, we began by asking their motivation for streaming. We asked how they set up their livestream, why do they set up the livestream like this, their attitude toward popularities, how do they foster the community and the challenges they have met. We asked whether the streamers have watched SWM livestream. We also asked the streamers who have watched SWM livestream the questions for viewers. For the viewers, we ask them the motivation for watching SWM livestream, what kind of streamers are the most attractive to them and their feedback of the streamers’ livestream practice. We sought to understand what kinds of interactions interviewees had with others on-stream and off-stream, within and beyond Bilibili, and what their intentions and concerns were. We then ask what the participants wished to improved of the streamers’ practice or the livestream platform. The interviews were conducted through Wechat Call. Each interview was recorded and lasted from forty-five minutes to two hours.

Data from the interviews were first transcribed and then analyzed using open coding to identify preliminary themes. We noticed the emergence of themes centered on self-regulation learning, companionship, and sense of belonging during the initial inductive coding process. In the second deductive coding process, we coded the interview content again with sensitivity to SRL and SOC. We paid special attention to the interviewees' practice, emotional experience and the influence of watching SWM livestream on learning behavior.

\subsection{Basic information of the participants}
\begin{table}[htb!]
	\centering
	\caption{Basic information of Streamers. (U.N.G.E.E. = Unified National Graduate Entrance Examination) }
	\begin{tabular}{m{2em}<{\centering}m{4em}<{\centering} m{3em}<{\centering }m{4em}<{\centering}m{4em}<{\centering}m{6em}<{\centering}m{6em}<{\centering}m{5em}<{\centering}}
		\toprule
		No.   & Gender & Age & Occupation & \makecell{Fans\\number} & Study Content & \makecell[l]{Livestream\\ experience} & \makecell{Watch SWM\\ livestream\\(Y/N)} \\
		\midrule
		S1    & Female & 23 & Unemployed  & 1639 & U.N.G.E.E. & 4 month & Y \\
		S2    & Female & 21 & Unemployed  & 19k   & U.N.G.E.E. & 5 month & Y \\
		S3    & Female & 21 & Student & 9336 & \makecell{Daily study\\ for school exam} & 2 years & Y \\
		S4    & Male  & 24    & Unemployed  & 10k   & U.N.G.E.E. & 6 month & Y \\
		S5    & Male  & 19 & Student & 129 & Programing & 2 month & N \\
		S6    & Female & 24 & Unemployed  & 2368 & U.N.G.E.E. & 6 month & Y \\
		S7    & Female & 24 & Unemployed  & 552 & U.N.G.E.E. & 3 month & Y \\
		S8    & Female & 24 & Unemployed  & 7898 & IELTS & 1year & Y \\
		S9   & Female & Privacy & Accountant & 225 & \makecell[l]{English learning \\ \& CPA \& MPAcc} & 7 month & Y \\
		S10   & Female & 22 & Student & 58k & U.N.G.E.E & 1 month & Y \\
		S11   & Male  & 26 & Unemployed & 477K & U.N.G.E.E & 1.5 year & Y \\
		S12 & Female & 23 & Student & 7537 & \makecell{Daily study\\ tasks} & 3 month & N\\
		\bottomrule
	\end{tabular}%
	\label{tab2}%
\end{table}%

\begin{table}[htb!]
	\centering
	\caption{Basic information of viewers.}
	\begin{tabular}{m{2em}<{\centering}m{4.055em}<{\centering}m{2em}<{\centering}m{5.055em}<{\centering}m{6.665em}<{\centering}m{10.22em}<{\centering}}
		\toprule
		No    & Gander & Age & Occupation & \makecell{Experience\\ of watching\\SWM livestream} & Study goals \\
		\midrule
		V1    & Male  & 21 & Student & 6 mon  & U.N.G.E.E. \\
		V2    & Male  & 20 & Student & 2 mon  & U.N.G.E.E. \\
		V3    & Male  & 22 & Student & 3 mon  & U.N.G.E.E. \\
		V4    & Female & 23 & Student & 2 mon  & Daily learning tasks \\
		V5    & Male  & 22 & Student & 8 mon  & School exam \\
		V6    & Female & 23 & Student & 5 mon  & Daily learning task \\
		V7    & Female & 24 & Student & 6 mon  & U.N.G.E.E. \\
		V8    & Male  & 24 & Student & 3 mon  & Daily learning tasks \\
		V9    & Male  & 20 & Student & 2 mon  & Daily learning tasks \\
		V10   & Female & 19 & Student & 4 mon  & U.N.G.E.E. \\
		\bottomrule
	\end{tabular}%
	\label{tab3}%
\end{table}%
The background information of the streamers in our interview is shown in Table \ref{tab1}. (we use S for streamers and V for viewers). Some of them are students (S3,5,10,12). Some of them just graduated from university, but they didn’t want to work. So, they put all their time and energy into study and prepare for the graduate entrance examination (S1, 2, 4, 6, 7, 8, 11). There was also streamers who have worked for some years but quit her job and want to obtain a vocational qualification or pursue a master degree for better job opportunities (S9). Most of them livestream for more than 10 hour a day(S1, 2, 4, 6, 7, 8, 9, 10, 11). Those who streams daily study tasks stream for more than 3 hours a day (S3, 5, 12). Most of the streamers we have interviewed are female (3 male, 9 female). The fans number of streamers vary from 225 to 477K. All the viewers we reached were college students. 

The basic information of the viewers participated in our interview is shown in Table \ref{tab2}. And there are 6 males and 4 females. We found that the viewers watch SWM livestream when they are preparing for examinations.
\section{Findings}
We summarized and sorted out the activities of SWM livestream on Bilibili, and then described their effects on SRL and SOC combined with observation and interviews. We found that SWM livestreams on Bilibili often have a detailed title, including information of their study goals and study schedules. The streamers disclose their personal information and have strict rules and punishment to limit the viewers’ interactions. It is worth noting that the checking in and study process report session of SWM livestream help the viewer’s SRL and keep them engaged. Streamers expanded the SRL community through QQ or WeChat groups and virtual study room tools beyond Bilibili. We uncovered the streamer's motivation that they produced livestream to look for supervision, find like-minded partners, and help and company others.
\subsection{An overview of SWM livestream}
In our observation and data analysis, the average number of SWM livestreams is 1558 per day. The average viewer number is 435. Some educational institutions use Bilibili livestream for teaching and they also category their livestream in this directory. And some streamers didn’t disclose any information about their goals. Excluding these livestreams, there are 652 livestreams remain and we classified the study goals of these streamers. The number and percentage of this goals are shown in table \ref{tab1}.
\begin{table}[htb!]
	\centering
	\caption{The types of examination the streamers are preparing.(CPA=Certified Public Accountant, MPAcc= Master of Professional Accounting)}
	\begin{tabular}{m{18.22em}ccc}
		\toprule
		Types  & Number & Percentage \\
		\midrule
		post-graduate entrance examination & 274   & 42\% \\
		Homework or exam at school (university) & 220   & 34\% \\
		college entrance examination & 59    & 9\% \\
		Upgraded from unified recruitment & 10    & 2\% \\
		Civil service examination, examination for staffing of government-affiliated institutions, CPA and MPAcc & 51    & 8\% \\
		National Judicial Examination & 24    & 4\% \\
		IELTS and TOEFL & 12    & 2\% \\
		Total & 652   &  \\
		\bottomrule
	\end{tabular}%
	\label{tab1}%
\end{table}%

According to Table\ref{tab1}, we found that streamers mainly study for entrance examination, job qualification and school examination. In China, entrance examination and job qualification are of great significance for people because they play a key role in future development. These examinations involve many subjects, require a high level of knowledge in a certain field, and are highly competitive because of the large number of participants. As a result, test-takers often need months or even years to prepare.
\subsection{The Practices and activities of SWM on Bilibili}
In this section, we summarized and categorized the activities of SWM livestream and showed the result in table 4. Some of the activities are consistent with the research on SWM Video \cite{lee2021personalizing}, watching SWM livestream give the viewers a chance to build a good learning atmosphere and urge themselves to study. In the following work, we mainly discuss the unique practices and activities presented in SWM livestream.
\subsubsection{Detailed self-introduction}
According to their study goals, streamers often introduce themselves in detail on their livestream screen, livestream introduction, moment or column (18/20). For example, a streamer preparing for the postgraduate entrance examination would introduce his education background (the university they graduated from, undergraduate major etc.), the university and major of the postgraduate entrance examination they are preparing for, and the reason for livestream, etc. Some streamers would also list the information and purchase links of stationery and livestream products they use in the introduction.

For streamers, disclosing all the information to the viewers at once can save their time and energy. They can avoid unnecessary questions and answers because many viewers are asking the same questions. Just as S6 described:

\textit{"I wanted to perfect the text on the livestream picture because sometimes the viewers would ask me questions such as 'What major do you apply for?'. Many viewers have asked the same question, but there was always somebody who didn't know that. So, I wrote down this information. I just accumulated little by little. When I thought it was perfect, I didn’t change it anymore."}

This information met viewer's need for information. The content of the exam goals and study experience in the introduction makes it easier for the viewers to find people with similar goals and to join a suitable and identifiable SRL community.

V5:\textit{"I read the streamer's introduction and choose streamers preparing the same examination with me because we are facing the same challenge. It's very reassuring to look at them."}

Some streamers even show their study method and experience in their self-introduction. Therefore, the introduction can serve as a reference for the viewers with similar experiences, inspiring viewers to reflect on their study methods and prompting them to make changes. S8 watched a SWM livestream for a long time. She was greatly influenced by the streamer. S8 was preparing for studying abroad, and English was a great challenge. S8:\textit{"The girl I subscribed to has been streaming her study footage for a long time, almost a year. It inspired me because some of our experiences were pretty similar. We all major in arts. I haven't passed CET-4 and CET-6, and my English is poor.  I preferred the quick way on the ADs before, but I found it doesn’t work well. It was because of the streamer, I realized it is not very realistic to improve my IELTS scores within three or five months as the mainstream IELTS classes ADs say. Then I made up my mind to do solid work and to study English step-by-step. She has a significant influence on me. "}  

Also, the detailed self-introduction can present a real and vivid image of the streamer and motivate the viewers. Some of the streamers' profiles described themselves as a persistent, hard-working person, which motivated and inspired the viewers. 

V10 explained that:  \textit{"The streamer's experience is interesting to me because he graduated from a junior college. That makes his experience of preparing for postgraduate entrance examination more inspirational. Because his knowledge base is too poor, but he can work so hard and stick to it for a longer time than me. I think that's what motivates me."}
\subsubsection{Rules and punishment to regulate on-stream social activities}
In prior research on livestream, streamers often constructed a welcoming personality, respond to a timely manner to the Danmakus during livestream and try to meet the interaction requirements of the viewer \cite{15}\cite{lee2018tip}\cite{23}. However, the streamers often inform the streamers that they don’t respond the Danmaku during the livestream because they want to concentrate on studying and don't want to be disturbed (S1,S2,S5,S6,S7,S8,S9,S10,S11). SWM livestream have fixed interaction sessions. Streamers declare these rules in advance and allow the viewers to participate spontaneously.

First of all, streamers use strict rules to regulate the viewer’s on-stream social activities (4/20). They show chatting manners in the introduction to avoid talking insult, negative or improper words (such as trolling others, advertising, distributing pirated materials, flooding the screen, etc.). For these negative comments, they would directly delete the words and block the viewers (S9, S4). This can help them prevent the viewers from being deceived or interrupted, and create a positive study environment. Like S11 explained:

\textit{"Because some viewers are very young, especially those students in high school, they may not be able to distinguish between right and wrong. So, I would kick out and block some many negative and aggressive words that affect the atmosphere."}

And S4 said:
\textit{"I directly block the rude remarks. And some Danmakus are no malicious but affect the atmosphere, I also block them. It is normal that there are trollers because our life is different. I just don't want to let it influence us."}
\subsubsection{Check-in and points query}
Chat-bot and widget that are only auxiliary interaction in prior research \cite{16} are widely used in SWM livestream as primary interaction. Check-in (9/20) is an activity that the viewers can send a specified Danmaku as a password, as an evidence to prove that they have come. For example, S11 specified \textit{"I can still learn"} as the password. Corresponding to the password, streamers use a widget to count the number of times each viewer has checked in. For example, a viewer can get 10 points each time they check in in S11' livestream. Thus, a viewer's score reflects the frequency they check in.To check their total points, the viewers can sent the comment \textit{“query”}. Viewers can get higher points by watching livestreams frequently and checking in more often, which means they are devoted fans. When the points reach a certain level, streamers can become the moderator. Similar to Twitch's study by Hamilton et al.\cite{23}, moderators can help streamers manage the livestream.

However, the main purpose of many viewers to check in is not to be a moderator, but to regulate their study behavior and form good study habits (V4, V5, V10).

As V5 said, \textit{"Persist in checking in can help me form a habit. This behavioral change and persistence seem to be more motivating."}

When many viewers participate in the check-in activity, they can sense each other's presence and feel a group accompanying and supporting them.

As described by V10:
\textit{"When you open the livestream, you can see many Danmakus saying “I can still learn”. It really works and can cheer you up, when you're really tired."}
\subsubsection{Spontaneous study progress report from the viewers}
The study progress report (6/20) is a spontaneous interaction of the viewers to send Danmakus about the subjects they are going to study, such as \textit{"I'll Hurry up and finish my math homework and try to finish science homework tonight."}, \textit{"Start to do physics paper"}, etc.

This interaction can enhance the mutual perception between the viewers and encourage the viewers to implement their study plan according to their planning. 
Just as V8 said, \textit{"I can study according to my plan. After reporting the study progress, I feel that my plan has become a kind of commitment, so I have to finish it on time.”}

V9:\textit{ "Seeing the specific subjects they are learning, I may have a more realistic feeling that we are studying together. I feel that they all working hard, and I should study hard, too."}
\subsubsection{Viewers pay for the title to get the privileges}
As mentioned above, the viewers can buy titles to get privileges in a streamer’s livestream. Some streamers also offer extra privilege for the “captains”  (1/20). S11’s "Captains" can befriend him, get into a “captain group” on WeChat, and get in the Tencent meeting (a zoom-like APP for multiperson online meeting) to livestream their study footage with S11.

Viewers regard the title as an investment on their education. Attracted by the personality of the streamers, they hope to be the streamer's close friend. Studying with such a peer model can strengthen their study motivation.

Just as V10 said: \textit{"Slowly, I have got used to the existence of the streamer. I just want to be friend with him, so I bought the “captain” title and befriended him on WeChat. At first, he was just a streamer, a stranger to me, but now I think we might be called friends. And I can join the “captain group”, and Tencent meeting. It’s a different opportunity to see the streamer in another livestream room. It’s like an investment in my education. I paid for the title and felt a better study atmosphere."}
\subsection{Activities Beyond Livestreams}
To better support the SRL community, streamers set up QQ or WeChat groups beyond Bilibili, multi-person virtual study room to enhance social interaction and provide more help on study. However, the heavy workload of streamers makes it a challenge to manage these groups.
\subsubsection{Private Fan Groups for SRL support}
Some streamers want to help the viewers and themselves, so they set up QQ or WeChat study support groups at the request of the viewers (10/20). The main activities in these groups are checking in, asking for information about the examination, Q\&As about examination questions, sharing study materials and reciting material together. S1 said: \textit{“There are a lot of people who are taking the examination of Master of Law, so I set up a WeChat group to share the material and the questions they can’t solve.”} 
S6 explained:\textit{ "At first, I didn't think of building groups because I didn't have many fans. Later I build a group because one day, a viewer asked if there was a group to study together. Others said they could not get up early to study. Then I thought, 'Let's set up a group. We can find someone to wake each other up and supervise each other's study.' ”}  
S10 said \textit{ "I set up a group to recite politics materials with them, where everyone can recite together and check each other's recitation."}

Such groups are usually accessible to anyone who knows the group ID, so members' social identities varied and are difficult to verify. Therefore, the streamers also have chatting manners in the group announcement to avoid negative words and advertisements. S11 said\textit{ "I will ban any content advertising, selling courses or unrelated to study. Once I find out, I will clean them out directly."}

Because streamers still need to focus on their study, they have no time to manage these groups. Most of these groups are managed by chat-bots. The chat-bot declare the chatting manners and say some encouraging words to urge the members to study. Streamers also delete the inappropriate words and clean out those members. Gradually, the groups become autonomous communities for the viewers. S4 said: \textit{"I don't manage my group and just let them chat in the group and clean out the trollers."} 

Some streamers have a group for the viewers with the special title mentioned above (1/20). The threshold of entering such group is higher, members are usually loyal fans of the streamer and have a closer relationship with the streamer. Therefore, such groups don't have strict rules and the members can talk about trivia in life. V10 said: \textit{"Our captain group is very funny. There is no limit. Every day we check in to learn and chat when we are tired. For instance, we show our take-out and cuisine in the group. The captain group is funny, but sometimes very quiet. Because everyone is studying and busy."}
\subsubsection{Why didn't set up a private fan group}
Many streamers didn't want to set up any private groups because they thought that SWM livestream is a personal behavior. They are happy that SWM livestream helped themselves and viewers study, but they are unwilling to let the viewers know more about themselves. Therefore, they didn’t build a group or share their SNS IDs. As S7 explained: \textit{"Once I wanted to have a group, but I pay attention to privacy. I think Bilibili is a public platform while WeChat and QQ are private things. I don’t want others to have an easy access to my QQ or WeChat ID."}

Setting private groups also means more social pressure and chances to be influenced by others. S8 said: \textit{" I want to keep my own pace and be myself. I don't want to change myself because of other people's opinions. I stream just for study and don’t want anything else."}

In addition, it takes time and energy to maintain group chats, and streamers can't get valuable information from group chat. S8 said:\textit{ "More people in the group mean you have to spend more time and energy to maintain it. You should check the message every day but can not learn much from it."} And S7 said:\textit{" I'm lazy, and if I build a fan group, I have to maintain it. that’s troublesome. I'm not that good at chatting, so I can't cheer up the atmosphere. It would be embarrassing if they are not active in the group. And I thought that a viewer may like many streamers and uploaders. It would be embarrassing if one day they don’t like me and people don’t say anything in the group. "}
\subsubsection{Multi-person virtual study room beyond Bilibili}
Some streamers set multi-person virtual study rooms (10/20) using multi-person meeting APP or study-support APPs to support several people studying together. Streamers organize multi-person study room sessions in two ways. One way is to stream the study footage with the viewers in a group meeting APP such as Tencent Meeting, where every one can livestream their study footage. The other is to set up collective productivity tracking groups in APPs such as Tomato ToDo, where members' hours of study can be recorded, ranked, and presented in a ranking list. These study groups give the members a better sense of studying together and strengthen the emotional connection. V10 said:\textit{“I can see their learning progress. And the members in the study room are familiar study partners so I can get a strong sense of being with them.”}

However, this kind of collective virtual study room can only accommodate a limited number of people. To motivate the members to study for longer time, and use the “space” of the “study room” efficiently, the streamers set minimum standards for how long members should study. The viewers who didn’t study for enough time would be cleared out of the group. S11:\textit{"Every night after the livestream, I will clear out those who doesn't study enough time. The purpose of this study rooms is to encourage them to study long enough every day."} S12:\textit{"Before 22:00 every day, they should study at least 30min, otherwise they will be automatically recognized by the system, and then cleared out.”}
\subsection{Streamers’ motivation}
Most streamers produced SWM livestream to help their study because they have great study pressure. Their demand mainly has two aspects. One is the lack of external supervision. The streamers use livestream to strengthen the supervision of SRL. The second is the lack of a study community with a sense of belonging and sense of identity. Streamers use SWM livestream to find more friends with similar educational experience and study goals to enhance the sense of community and get emotional support. Some streamers are not preparing for big examinations and don’t have much study pressure. They livestream to company others. They do it out of interest or as a kind of "We Media" practice.
\subsubsection{Seek supervision}
Streamers produce livestream to feel other’s presence as a kind of supervision. Then they can moderate and regulate their study behavior. 
Making a learning plan is often easy, but it can be challenging to implement, especially if it is a commitment to yourself alone. The SWM streamers puts their daily learning and work schedule or even live schedule in the live stream interface, and the viewers can monitor their own implementation, so that the stream has a sense of responsibility for the plan. S8 is preparing for IELTS and planning to study abroad. SWM makes her study more regularly. \textit{“Before I started streaming, I learned more freely. But once I start to livestream, I can see the number of viewers, especially when there are a relatively large number of people. I became more strict with myself. When I cannot recite words, I may become irritable. Or when I want to give up and do something else, or when I am distracted, I may think that so many people are watching me. If I don't act according to my original plan, or if I suddenly disappeared, I may have a feeling that what I said to others has not been achieved. There is a feeling of being supervised and urged by others. Because of this, I can study for a longer time. "}

\subsubsection{Find like-minded study partners}
For most streamers, the lack of emotional support and a sense of belonging is a great challenge in preparing for the exam. It can take months or even a year to prepare for a big examination, and learners desperately need others' understanding, companionship, and emotional support during the long preparation process.

Especially for those who have interdisciplinary U.N.G.E.E. or have different learning goals from their surrounding classmates, it is difficult for them to find students with the same goal. SWM livestream helped them find peers in a broader circle. And streamers can find like-minded sutdy partners to study together, accompany and encourage each other.

S2 majored in social science before and is preparing for her U.N.G.E.E. in computer science the second time. Her goal is to be accepted by a first-class university in China. It is difficult for her to find students with a common language in this and study together.\textit{ “My social circle is not so big, and the people in my circle seldom know computer science.And I am not a fresh graduate, there are few classmates around me. It’s not so easy to find study companions in daily life. But I'm lucky to find many candidates for computer science major. They helped me a lot.”} 

Study with these like-minded friends also help the streamers to reflect on their learning plan and behavior. Streamers can communicate with viewers about their study and know the learning progress of other examinees. S6 said:\textit{"when I study at home alone, I don't know other students’ learning study progress and what level was I at then. But I get many friends preparing the same examination, I can have a more objective understanding of my study through communication with others."} 

\subsubsection{Help and company others}

Some streamers do not have the pressure of examinations. They livestream not to for study progress but to share life, company others, or livestream as a we-media practice.

S3 has been studying abroad in the United States, and her academic performance has always been excellent. For her, doing the SWM livestream is a way to enrich her life, and to give the viewers a sense of intimacy and companionship. \textit{"I was inspired by a friend of mine. He was in New Zealand, and he often streams on YouTube. Watching him on livestream gives me a good feeling of studying with him. And I feel that he is really close to me. Later, I started the SWM livestream with him on YouTube, and moved to Bilibili at the request of domestic friends.”} S12 said:\textit{"I don't have any examination pressure. I stream my daily study tasks and want to gain more fans."}
\subsection{Challenges of SWM livestream}
\subsubsection{SWM livestream disappear when the streamers finish their examination} 
Streamers produce SWM livestream because of their study goals. This makes the SWM livestream not sustainable for a long time. Once the examination is over and the streamers achieve their goal, most the streamers would stop their livestream. Some even cleared their feeds and columns posted on Bilibili, and had their Bilibili accounts disappear entirely.

For streamers, SWM livestream is just a short journey with the viewers, after which they have to cut off all connections with them. S8 said:\textit{"I just take responsibility to accompany them through this journey. When the postgraduate entrance examination is over, and everyone has a good result, it’s time to stop it.”}

Some streamers believe that the fan base accumulated by SWM livestream can serve as a springboard for them to become streamers or uploaders in other areas in the future. S4 said:\textit{" The fans just means to be recognized by everyone as an SWM streamer and lay a foundation for the future transition to uploading other kinds of videos."}
\subsubsection{Streamers face questions about their motivations and appearance}
The streamers of SWM Livestream are faced with many doubts. On the one hand, these doubts come from the contradiction between the seriousness of study and the entertainment of livestream. There are often confused comments questioning the motives of streamers. Some visitors thought that livestream is a means of entertainment. SWM livestream is more like showing off themselves and  rather than contributing to their study. S2 said:\textit{"Some people look down on us. They think it is strange, it is useless and our grades will not get better. Everyone's study style is different, and those who can learn without any external force are indeed the best. However, it would be a breakthrough if those who are not so self-disciplined could surpass themselves with the help of livestream. Not everyone should compare with others."} S8 said:\textit{ "Some people would say, are you learning or are you deliberately attracting fans?"} S11 said some visitors doubt:\textit{"Are you livestream for money?"}

Faced with the pressure, streamers regulate their words and deeds. They avoid too many economic-related activities in their livestreams. The streamers extra priviledges for "captain" title we mentioned only lasts for a few months and then disappeared from the streamer’s studio because it is an obvious guidance to consume. S11 said:\textit{"Many people think that I am a fraud, and scolded me badly. So, I received an advertisement but did not do it."}

In SWM livestream practice in China, few streamers show their face in SWM livestream. The streamers who show their face in livestream get a lot of negative comments. Some criticised the streamers’ appearance. Some have malicious speculation about streamers’ motivation. Some viewers think that showing the face is a soft porn to get attention, especially female streamers. S12 said:\textit{ "I showed my face because I had no other shooting equipment, I had to use my mobile phone to stream. But after I showed my face, many people said 'that’s pornography', 'Why do you stream study and still have time to make up?'. That was very disturbing, and then I directly blocked all the comments containing this kind of keywords."}
\section{Discussion}
Our work reveals how novel activities of SWM livestream facilitated SRL community. In this section, we analyze how streamers regulate activities in SWM Livestream, how these activities support SRL, how SOC develops in livestream, and in turn, SOC promote the development of the community.
\subsection{The activities of SWM livestream help with SRL}
SRL consists of regulation of different dimensions such as cognition, motivation, and emotion. And according to Pintrich model, SRL is compounded by four phases: (1) Forethought, planning, and activation; (2) Monitoring;(3) Control; and (4) Reaction and reflection \cite{pintrich2000role}. We found that in the practice of SWM livestream, various activities support the SRL of streamers and viewers from different dimensions.

 From the cognitive dimension, SWM Livestream encourages members to actively reflect on their study situation and study plan, then adjust and change their plan and behaviors. The streamers’ schedule, the progress reported session, and the chatting about study helped them to situate their study progress in a wider range, and make further adjustments. In addition, the checking-in and progress report sessions regulate members' study behaviors and stimulate their incentive to form regular and continuous study behaviors.

Motivational and emotional, streamers control and limit the topics and action in activities to effectively concentrate members on study-related topics and avoid distraction. The experience and encouragement from others provide members with emotional support. Consistent with some social psychological research, streamers going through similar experiences bring strength and comfort to the viewers and bring some certainty to an uncertain learning path \cite{12}\cite{18}\cite{26}\cite{32}\cite{45}. Therefore, the members’ stress can be relived, and they can have confidence and courage to persist in the long-term study.
\subsection{SOC developed in SWM livestream}
We noticed that most of the streamers and viewers participated in SWM livestream because they were faced with the stress of exams, especially big national exams, which required long-term preparation. They need a community to support them and gain a sense of being understood and a sense of belonging. SOC plays a crucial role in SWM Livestream. It provides members with emotional sustenance and support, enabling them to keep study motivation and persistence in the preparation process of examination, which lasts for months or even years. The basic elements of SOC are reflected in SWM: membership, influence, integration and fulfillment of needs and shared emotional connection.

The establishment of membership in SOC includes several elements: boundaries, emotional safety, sense of belonging and identification, personal investment and a common symbol system \cite{mcmillan1986sense}. All of these elements are represented in SWM livestream. In terms of membership, the streamers set the boundaries by stating like “no study, no enter”, thereby defining the membership criteria and provide security that protects group intimacy. As for the sense of belonging and identity, streamers control their information disclosure such as showing their educational background and experience to present a diligent, tenacious, friendly, independent and individualized personality, which overlap with the viewer’s experience and is the viewers' desirable personality. Therefore, they can feel that they have found the community fits them. Viewers regard tipping and title purchase as a kind of investment, strengthening the sense of membership. The shared emotional connection is shaped through on-stream and off-stream activities. The checking in and progress report session strengthened the connection by promoting viewers' social presence, which increased the number of members and community size they perceived. The limitation of topics and punishment mechanisms enhances emotional security and gives the community a more positive experience, enhancing the bond and facilitated cohesion.

As mentioned in the above section, SWM livestream support SRL, which fulfill the members' need on study. In turn, consistent with prior research, SOC facilitates the maintenance and development of online learning community \cite{gilbert1998building}. SOC ensures the retention of members in SRL community and enables SWM livestream to have a long-term impact on SRL. V10 was a freshman in college, she has watched a streamer's livestream for more than one year, and she said:\textit{ "I think I am even used to the existence of this livestream. If one day you suddenly tell me this livestream was disappeared. What would I do? I may be heartbroken, so I hope that I were a senior and take the entrance examination for graduate school, I would still be able to watch this livestream."}

\subsection{The regulation of SWM activities}
In prior research, streamers in knowledge-sharing streams found it challenging to keep their viewers engaged while focusing on more performative activities \cite{34}\cite{16}\cite{15}. SWM livestream practice inspired us how to reduce the interactive pressure of streamer. Although streamers have no time to take care of on-stream real-time activities, they have strategies to regulate the activities. Whether it is on-stream or off-stream, streamers often do not directly interact with the viewers, but take advantage of checking in, automatic responds and the viewers’Danmaku sideshows to ensure viewers' engagement and served the goals of SRL. They respond to private message in their spare time and only add a few people's SNS and become real friends with them. S6 said:\textit{"I respond to the private massages when I want to take a break, and it only takes me about an hour to do that every day."} S7 said:\textit{" I got on well with a few subscribers through Bilibili private messages. We encourage and cheer up each other, and then I befriend in WeChat. But I don’t befriend many people, so far just have added three or four good friends." }

Streamers adopted strict rules to limit the topics and content, making it study-related and can produce a positive study atmosphere. Refer to the classification of task-related and social activities in prior research. We found that SWM streamers encourage Greetings and Asking questions, Social support, and Shared understanding while avoiding Social resistance and Loss of Shared understanding. This activity strategy can maintain a positive atmosphere and reduce the interaction burden of the streamers. However, this also makes the chatting content of Danmaku and in the study-support group relatively simple and members' participation not active enough.

Streamers expand the SRL community beyond Bilibili by establishing private fans groups and multi-person virtual study rooms (multi-person livestream or multi-person learning progress record). These communities also have topic limits and study incentives to support members' SRL. Multi-person self-study is an extension of SWM livestream. It makes up for the shortage that only one person's study footage can be seen and monitored, and the study progress among members can be tracked in real-time. Although multi-person livestream is a better way of mutual supervision, it has higher requirements on the equipment, network, places, and the relationship between the members. So, more streamers use multi-person process record groups to provide members with the opportunity to study together.

By studying SWM livestream on Bilibili and its expansion activities beyond Bilibili, we found that SWM Livestream is an intermediary and bridge for learners to get to know each other. Public communication in SWM livestream or the group is stereotyped and limited. Deeper communication and relationship between the individuals are further developed after they befriend each other in SNS.
\section{LIMITATIONS AND RECOMMENDATIONS FOR FUTURE}
This paper reports findings from observation, and interview of streamers and viewers of SWM livestream on Bilibili. As a result, our study has several limitations.

First, the sampling strategy (purposive sampling and snow-ball sampling) used may produce biased samples. We call for caution in our results; Second, in our observation, we found that most of the streamers and viewers were students or recent graduates. Therefore, other ages and occupations are not presented much in this paper; Third, the only people who can be found are those who continue to stream or watch SWM livestream, but we can't find those who may have tried but gave up, so there may be a survivor effect. There are more females among the streamers who accept the interview, so we suppose to explore the gander difference of SWM livestream streamers. More systematic research in the future could avoid these limitations. Besides, a quantitative study of a more diverse and larger population would be a valuable complement to this qualitative, exploratory study. An extensive survey of SWM livestream streamer and viewer will help confirm and further develop our findings.
\section{CONCLUSIONS}
In this paper, we adopt observation and interview to better understand the practice and activities in SWM livestream and streamers motivation. We further understand how activities of SWM livestream support CSCSRL from cognitive dimension and SOC from the socioemotional dimension. We also identified the challenges of SWM livestream. We uncovered that streamers produced SWM livestream to seek supervision, find like-minded partners, and help and company others. The streamers practice of using rules, punishment, chat-bot, and automatic response effectively regulated the task-related and non-task-related activities and alleviated the streamer’s interaction pressure. The activities of SWM livestream helped members to concentrate on study, and also helped members reflect on their study plan and behaviors and regulate their study behaviors. In SWM livestream, like-minded members gathered together and constructed a community for SRL, where the SOC developed and prospered. Our findings of SWM livestream suggested opportunities to construct CSCSRL community for lonely learners working on a long-term study.

\appendix

\bibliographystyle{ACM-Reference-Format}
\bibliography{reference}


\begin{thebibliography}{47}


\ifx \showCODEN    \undefined \def \showCODEN     #1{\unskip}     \fi
\ifx \showDOI      \undefined \def \showDOI       #1{#1}\fi
\ifx \showISBNx    \undefined \def \showISBNx     #1{\unskip}     \fi
\ifx \showISBNxiii \undefined \def \showISBNxiii  #1{\unskip}     \fi
\ifx \showISSN     \undefined \def \showISSN      #1{\unskip}     \fi
\ifx \showLCCN     \undefined \def \showLCCN      #1{\unskip}     \fi
\ifx \shownote     \undefined \def \shownote      #1{#1}          \fi
\ifx \showarticletitle \undefined \def \showarticletitle #1{#1}   \fi
\ifx \showURL      \undefined \def \showURL       {\relax}        \fi
\providecommand\bibfield[2]{#2}
\providecommand\bibinfo[2]{#2}
\providecommand\natexlab[1]{#1}
\providecommand\showeprint[2][]{arXiv:#2}

\bibitem[\protect\citeauthoryear{Abedin, Daneshgar, and D'Ambra}{Abedin
  et~al\mbox{.}}{2012}]%
        {abedin2012nontask}
\bibfield{author}{\bibinfo{person}{Babak Abedin}, \bibinfo{person}{Farhad
  Daneshgar}, {and} \bibinfo{person}{John D'Ambra}.}
  \bibinfo{year}{2012}\natexlab{}.
\newblock \showarticletitle{Do nontask interactions matter? The relationship
  between nontask sociability of computer supported collaborative learning and
  learning outcomes}.
\newblock \bibinfo{journal}{\emph{British Journal of Educational Technology}}
  \bibinfo{volume}{43}, \bibinfo{number}{3} (\bibinfo{year}{2012}),
  \bibinfo{pages}{385--397}.
\newblock


\bibitem[\protect\citeauthoryear{app.mi.com}{app.mi.com}{2021}]%
        {costudy}
\bibfield{author}{\bibinfo{person}{app.mi.com}.}
  \bibinfo{year}{2021}\natexlab{}.
\newblock \bibinfo{booktitle}{\emph{Costudy}}.
\newblock
\urldef\tempurl%
\url{https://app.mi.com/details?id=com.costudy.costudy}
\showURL{%
Retrieved July 13, 2021 from \tempurl}


\bibitem[\protect\citeauthoryear{baike.baidu}{baike.baidu}{2021}]%
        {bilibili}
\bibfield{author}{\bibinfo{person}{baike.baidu}.}
  \bibinfo{year}{2021}\natexlab{}.
\newblock \bibinfo{booktitle}{\emph{Bilibili}}.
\newblock
\urldef\tempurl%
\url{https://baike.baidu.com/item/bilibili}
\showURL{%
Retrieved July 13, 2021 from \tempurl}


\bibitem[\protect\citeauthoryear{Bannert and Reimann}{Bannert and
  Reimann}{2012}]%
        {bannert2012supporting}
\bibfield{author}{\bibinfo{person}{Maria Bannert} {and} \bibinfo{person}{Peter
  Reimann}.} \bibinfo{year}{2012}\natexlab{}.
\newblock \showarticletitle{Supporting self-regulated hypermedia learning
  through prompts}.
\newblock \bibinfo{journal}{\emph{Instructional Science}} \bibinfo{volume}{40},
  \bibinfo{number}{1} (\bibinfo{year}{2012}), \bibinfo{pages}{193--211}.
\newblock


\bibitem[\protect\citeauthoryear{CCTV.com}{CCTV.com}{2019}]%
        {1}
\bibfield{author}{\bibinfo{person}{CCTV.com}.} \bibinfo{year}{2019}\natexlab{}.
\newblock \bibinfo{booktitle}{\emph{Do you know that this generation of young
  people would love to study on Bilibili}}.
\newblock
\urldef\tempurl%
\url{http://news.cctv.com/2019/04/17/ARTIkdxgldxCuSmVdTOimrAw190417.shtml}
\showURL{%
Retrieved April 17, 2019 from \tempurl}


\bibitem[\protect\citeauthoryear{Chen, Freeman, and Balakrishnan}{Chen
  et~al\mbox{.}}{2019}]%
        {8}
\bibfield{author}{\bibinfo{person}{Di Chen}, \bibinfo{person}{Dustin Freeman},
  {and} \bibinfo{person}{Ravin Balakrishnan}.} \bibinfo{year}{2019}\natexlab{}.
\newblock \showarticletitle{Integrating Multimedia Tools to Enrich Interactions
  in Live Streaming for Language Learning}. In
  \bibinfo{booktitle}{\emph{Proceedings of the 2019 CHI Conference on Human
  Factors in Computing Systems}}. \bibinfo{pages}{1--14}.
\newblock


\bibitem[\protect\citeauthoryear{Chen, Lasecki, and Dong}{Chen
  et~al\mbox{.}}{2021}]%
        {chen2021towards}
\bibfield{author}{\bibinfo{person}{Yan Chen}, \bibinfo{person}{Walter~S
  Lasecki}, {and} \bibinfo{person}{Tao Dong}.} \bibinfo{year}{2021}\natexlab{}.
\newblock \showarticletitle{Towards Supporting Programming Education at Scale
  via Live Streaming}.
\newblock \bibinfo{journal}{\emph{Proceedings of the ACM on Human-Computer
  Interaction}} \bibinfo{volume}{4}, \bibinfo{number}{CSCW3}
  (\bibinfo{year}{2021}), \bibinfo{pages}{1--19}.
\newblock


\bibitem[\protect\citeauthoryear{Cirillo}{Cirillo}{2006}]%
        {cirillo2006pomodoro}
\bibfield{author}{\bibinfo{person}{Francesco Cirillo}.}
  \bibinfo{year}{2006}\natexlab{}.
\newblock \showarticletitle{The pomodoro technique (the pomodoro)}.
\newblock \bibinfo{journal}{\emph{Agile Processes in Software Engineering and}}
  \bibinfo{volume}{54}, \bibinfo{number}{2} (\bibinfo{year}{2006}),
  \bibinfo{pages}{35}.
\newblock


\bibitem[\protect\citeauthoryear{Dickerson and Kemeny}{Dickerson and
  Kemeny}{2004}]%
        {12}
\bibfield{author}{\bibinfo{person}{Sally~S Dickerson} {and}
  \bibinfo{person}{Margaret~E Kemeny}.} \bibinfo{year}{2004}\natexlab{}.
\newblock \showarticletitle{Acute stressors and cortisol responses: a
  theoretical integration and synthesis of laboratory research}.
\newblock \bibinfo{journal}{\emph{Psychological bulletin}}
  \bibinfo{volume}{130}, \bibinfo{number}{3} (\bibinfo{year}{2004}),
  \bibinfo{pages}{355}.
\newblock
\showISSN{1939-1455}


\bibitem[\protect\citeauthoryear{Faas, Dombrowski, Young, and Miller}{Faas
  et~al\mbox{.}}{2018}]%
        {15}
\bibfield{author}{\bibinfo{person}{Travis Faas}, \bibinfo{person}{Lynn
  Dombrowski}, \bibinfo{person}{Alyson Young}, {and} \bibinfo{person}{Andrew~D
  Miller}.} \bibinfo{year}{2018}\natexlab{}.
\newblock \showarticletitle{Watch me code: Programming mentorship communities
  on twitch. tv}.
\newblock \bibinfo{journal}{\emph{Proceedings of the ACM on Human-Computer
  Interaction}} \bibinfo{volume}{2}, \bibinfo{number}{CSCW}
  (\bibinfo{year}{2018}), \bibinfo{pages}{1--18}.
\newblock
\showISSN{2573-0142}


\bibitem[\protect\citeauthoryear{forest.cc}{forest.cc}{2020}]%
        {forest}
\bibfield{author}{\bibinfo{person}{forest.cc}.}
  \bibinfo{year}{2020}\natexlab{}.
\newblock \bibinfo{booktitle}{\emph{Forest}}.
\newblock
\urldef\tempurl%
\url{https://www.forestapp.cc}
\showURL{%
Retrieved July 16, 2020 from \tempurl}


\bibitem[\protect\citeauthoryear{fqtodo.cn}{fqtodo.cn}{2021}]%
        {fqtodo}
\bibfield{author}{\bibinfo{person}{fqtodo.cn}.}
  \bibinfo{year}{2021}\natexlab{}.
\newblock \bibinfo{booktitle}{\emph{FanQieToDo official website}}.
\newblock
\urldef\tempurl%
\url{https://fqtodo.cn/}
\showURL{%
Retrieved June 16, 2021 from \tempurl}


\bibitem[\protect\citeauthoryear{Fraser, Kim, Thornsberry, Klemmer, and
  Dontcheva}{Fraser et~al\mbox{.}}{2019}]%
        {16}
\bibfield{author}{\bibinfo{person}{C~Ailie Fraser}, \bibinfo{person}{Joy~O
  Kim}, \bibinfo{person}{Alison Thornsberry}, \bibinfo{person}{Scott Klemmer},
  {and} \bibinfo{person}{Mira Dontcheva}.} \bibinfo{year}{2019}\natexlab{}.
\newblock \bibinfo{booktitle}{\emph{Sharing the studio: How creative
  livestreaming can inspire, educate, and engage}}.
\newblock \bibinfo{pages}{144--155}.
\newblock


\bibitem[\protect\citeauthoryear{freedom.to}{freedom.to}{2020}]%
        {freedom}
\bibfield{author}{\bibinfo{person}{freedom.to}.}
  \bibinfo{year}{2020}\natexlab{}.
\newblock \bibinfo{booktitle}{\emph{Freedom}}.
\newblock
\urldef\tempurl%
\url{https://freedom.to}
\showURL{%
Retrieved July 16, 2020 from \tempurl}


\bibitem[\protect\citeauthoryear{Frijda and Mesquita}{Frijda and
  Mesquita}{1994}]%
        {18}
\bibfield{author}{\bibinfo{person}{Nico~H Frijda} {and} \bibinfo{person}{Batja
  Mesquita}.} \bibinfo{year}{1994}\natexlab{}.
\newblock \showarticletitle{The social roles and functions of emotions}.
\newblock  (\bibinfo{year}{1994}).
\newblock
\showISSN{1557982244}


\bibitem[\protect\citeauthoryear{Gilbert and Moore}{Gilbert and Moore}{1998}]%
        {gilbert1998building}
\bibfield{author}{\bibinfo{person}{Larry Gilbert} {and}
  \bibinfo{person}{David~R Moore}.} \bibinfo{year}{1998}\natexlab{}.
\newblock \showarticletitle{Building interactivity into Web courses: Tools for
  social and instructional interactions}.
\newblock \bibinfo{journal}{\emph{Educational Technology}}
  \bibinfo{volume}{38}, \bibinfo{number}{3} (\bibinfo{year}{1998}),
  \bibinfo{pages}{29--35}.
\newblock


\bibitem[\protect\citeauthoryear{girl}{girl}{2021}]%
        {2}
\bibfield{author}{\bibinfo{person}{Lofi girl}.}
  \bibinfo{year}{2021}\natexlab{}.
\newblock \bibinfo{booktitle}{\emph{lofi hip hop radio - beats to relax/study
  to}}.
\newblock
\urldef\tempurl%
\url{https://www.youtube.com/watch?v=5qap5aO4i9A}
\showURL{%
\tempurl}


\bibitem[\protect\citeauthoryear{Hadwin, Oshige, Gress, and Winne}{Hadwin
  et~al\mbox{.}}{2010}]%
        {hadwin2010innovative}
\bibfield{author}{\bibinfo{person}{Allyson~F Hadwin}, \bibinfo{person}{Mika
  Oshige}, \bibinfo{person}{Carmen~LZ Gress}, {and} \bibinfo{person}{Philip~H
  Winne}.} \bibinfo{year}{2010}\natexlab{}.
\newblock \showarticletitle{Innovative ways for using gStudy to orchestrate and
  research social aspects of self-regulated learning}.
\newblock \bibinfo{journal}{\emph{Computers in Human behavior}}
  \bibinfo{volume}{26}, \bibinfo{number}{5} (\bibinfo{year}{2010}),
  \bibinfo{pages}{794--805}.
\newblock


\bibitem[\protect\citeauthoryear{Hamilton, Garretson, and Kerne}{Hamilton
  et~al\mbox{.}}{[n.d.]}]%
        {23}
\bibfield{author}{\bibinfo{person}{William~A Hamilton}, \bibinfo{person}{Oliver
  Garretson}, {and} \bibinfo{person}{Andruid Kerne}.}
  \bibinfo{year}{[n.d.]}\natexlab{}.
\newblock \showarticletitle{Streaming on twitch: fostering participatory
  communities of play within live mixed media}. In
  \bibinfo{booktitle}{\emph{Proceedings of the SIGCHI conference on human
  factors in computing systems}}. \bibinfo{pages}{1315--1324}.
\newblock


\bibitem[\protect\citeauthoryear{Hrastinski}{Hrastinski}{2008}]%
        {hrastinski2008asynchronous}
\bibfield{author}{\bibinfo{person}{Stefan Hrastinski}.}
  \bibinfo{year}{2008}\natexlab{}.
\newblock \showarticletitle{Asynchronous and synchronous e-learning}.
\newblock \bibinfo{journal}{\emph{Educause quarterly}} \bibinfo{volume}{31},
  \bibinfo{number}{4} (\bibinfo{year}{2008}), \bibinfo{pages}{51--55}.
\newblock


\bibitem[\protect\citeauthoryear{J{\"a}rvel{\"a}, Kirschner, Panadero,
  Malmberg, Phielix, Jaspers, Koivuniemi, and J{\"a}rvenoja}{J{\"a}rvel{\"a}
  et~al\mbox{.}}{2015}]%
        {jarvela2015enhancing}
\bibfield{author}{\bibinfo{person}{Sanna J{\"a}rvel{\"a}},
  \bibinfo{person}{Paul~A Kirschner}, \bibinfo{person}{Ernesto Panadero},
  \bibinfo{person}{Jonna Malmberg}, \bibinfo{person}{Chris Phielix},
  \bibinfo{person}{Jos Jaspers}, \bibinfo{person}{Marika Koivuniemi}, {and}
  \bibinfo{person}{Hanna J{\"a}rvenoja}.} \bibinfo{year}{2015}\natexlab{}.
\newblock \showarticletitle{Enhancing socially shared regulation in
  collaborative learning groups: Designing for CSCL regulation tools}.
\newblock \bibinfo{journal}{\emph{Educational Technology Research and
  Development}} \bibinfo{volume}{63}, \bibinfo{number}{1}
  (\bibinfo{year}{2015}), \bibinfo{pages}{125--142}.
\newblock


\bibitem[\protect\citeauthoryear{Jia, Shen, Chen, Li, and Iosup}{Jia
  et~al\mbox{.}}{2017}]%
        {jia2017analysis}
\bibfield{author}{\bibinfo{person}{Adele~Lu Jia}, \bibinfo{person}{Siqi Shen},
  \bibinfo{person}{Shengling Chen}, \bibinfo{person}{Dongsheng Li}, {and}
  \bibinfo{person}{Alexandru Iosup}.} \bibinfo{year}{2017}\natexlab{}.
\newblock \showarticletitle{An analysis on a YouTube-like UGC site with
  enhanced social features}. In \bibinfo{booktitle}{\emph{Proceedings of the
  26th International Conference on World Wide Web Companion}}.
  \bibinfo{pages}{1477--1483}.
\newblock


\bibitem[\protect\citeauthoryear{Ju-young}{Ju-young}{2018}]%
        {3}
\bibfield{author}{\bibinfo{person}{Park Ju-young}.}
  \bibinfo{year}{2018}\natexlab{}.
\newblock \bibinfo{booktitle}{\emph{Why do Koreans watch others studying alone
  on YouTube?}}
\newblock
\urldef\tempurl%
\url{http://www.koreaherald.com/view.php?ud=20180812000201&ACE_SEARCH=1}
\showURL{%
Retrieved July 14,2021 from \tempurl}


\bibitem[\protect\citeauthoryear{Keltner and Haidt}{Keltner and Haidt}{2001}]%
        {26}
\bibfield{author}{\bibinfo{person}{Dacher Keltner} {and}
  \bibinfo{person}{Jonathan Haidt}.} \bibinfo{year}{2001}\natexlab{}.
\newblock \showarticletitle{Social functions of emotions}.
\newblock  (\bibinfo{year}{2001}).
\newblock
\showISSN{157230622X}


\bibitem[\protect\citeauthoryear{Kim, Jeon, Choe, Lee, Kim, and Seo}{Kim
  et~al\mbox{.}}{2016}]%
        {kim2016timeaware}
\bibfield{author}{\bibinfo{person}{Young-Ho Kim}, \bibinfo{person}{Jae~Ho
  Jeon}, \bibinfo{person}{Eun~Kyoung Choe}, \bibinfo{person}{Bongshin Lee},
  \bibinfo{person}{KwonHyun Kim}, {and} \bibinfo{person}{Jinwook Seo}.}
  \bibinfo{year}{2016}\natexlab{}.
\newblock \showarticletitle{TimeAware: Leveraging framing effects to enhance
  personal productivity}. In \bibinfo{booktitle}{\emph{Proceedings of the 2016
  CHI Conference on Human Factors in Computing Systems}}.
  \bibinfo{pages}{272--283}.
\newblock


\bibitem[\protect\citeauthoryear{Kreijns, Kirschner, and Vermeulen}{Kreijns
  et~al\mbox{.}}{2013}]%
        {kreijns2013social}
\bibfield{author}{\bibinfo{person}{Karel Kreijns}, \bibinfo{person}{Paul~A
  Kirschner}, {and} \bibinfo{person}{Marjan Vermeulen}.}
  \bibinfo{year}{2013}\natexlab{}.
\newblock \showarticletitle{Social aspects of CSCL environments: A research
  framework}.
\newblock \bibinfo{journal}{\emph{Educational Psychologist}}
  \bibinfo{volume}{48}, \bibinfo{number}{4} (\bibinfo{year}{2013}),
  \bibinfo{pages}{229--242}.
\newblock


\bibitem[\protect\citeauthoryear{Lee, Chung, Song, Chang, and Kim}{Lee
  et~al\mbox{.}}{2021}]%
        {lee2021personalizing}
\bibfield{author}{\bibinfo{person}{Yoonjoo Lee}, \bibinfo{person}{John
  Joon~Young Chung}, \bibinfo{person}{Jean~Y Song}, \bibinfo{person}{Minsuk
  Chang}, {and} \bibinfo{person}{Juho Kim}.} \bibinfo{year}{2021}\natexlab{}.
\newblock \showarticletitle{Personalizing Ambience and Illusionary Presence:
  How People Use “Study with me” Videos to Create Effective Studying
  Environments}. In \bibinfo{booktitle}{\emph{Proceedings of the 2021 CHI
  Conference on Human Factors in Computing Systems}}. \bibinfo{pages}{1--13}.
\newblock


\bibitem[\protect\citeauthoryear{Lee, Yen, Chiu, King, and Fu}{Lee
  et~al\mbox{.}}{2018}]%
        {lee2018tip}
\bibfield{author}{\bibinfo{person}{Yi-Chieh Lee}, \bibinfo{person}{Chi-Hsien
  Yen}, \bibinfo{person}{Po-Tsung Chiu}, \bibinfo{person}{Jung-Tai King}, {and}
  \bibinfo{person}{Wai-Tat Fu}.} \bibinfo{year}{2018}\natexlab{}.
\newblock \showarticletitle{Tip me! Tipping is changing social interactions on
  live streams in China}. In \bibinfo{booktitle}{\emph{Extended Abstracts of
  the 2018 CHI Conference on Human Factors in Computing Systems}}.
  \bibinfo{pages}{1--6}.
\newblock


\bibitem[\protect\citeauthoryear{Locke and Horowitz}{Locke and
  Horowitz}{1990}]%
        {32}
\bibfield{author}{\bibinfo{person}{Kenneth~D Locke} {and}
  \bibinfo{person}{Leonard~M Horowitz}.} \bibinfo{year}{1990}\natexlab{}.
\newblock \showarticletitle{Satisfaction in interpersonal interactions as a
  function of similarity in level of dysphoria}.
\newblock \bibinfo{journal}{\emph{Journal of personality and social
  psychology}} \bibinfo{volume}{58}, \bibinfo{number}{5}
  (\bibinfo{year}{1990}), \bibinfo{pages}{823}.
\newblock
\showISSN{1939-1315}


\bibitem[\protect\citeauthoryear{Lu, Annett, Fan, and Wigdor}{Lu
  et~al\mbox{.}}{[n.d.]}]%
        {34}
\bibfield{author}{\bibinfo{person}{Zhicong Lu}, \bibinfo{person}{Michelle
  Annett}, \bibinfo{person}{Mingming Fan}, {and} \bibinfo{person}{Daniel
  Wigdor}.} \bibinfo{year}{[n.d.]}\natexlab{}.
\newblock \showarticletitle{" I feel it is my responsibility to stream"
  Streaming and Engaging with Intangible Cultural Heritage through
  Livestreaming}. In \bibinfo{booktitle}{\emph{Proceedings of the 2019 CHI
  Conference on Human Factors in Computing Systems}}. \bibinfo{pages}{1--14}.
\newblock


\bibitem[\protect\citeauthoryear{Lu, Annett, and Wigdor}{Lu
  et~al\mbox{.}}{2019}]%
        {35}
\bibfield{author}{\bibinfo{person}{Zhicong Lu}, \bibinfo{person}{Michelle
  Annett}, {and} \bibinfo{person}{Daniel Wigdor}.}
  \bibinfo{year}{2019}\natexlab{}.
\newblock \showarticletitle{Vicariously experiencing it all without going
  outside: A study of outdoor livestreaming in China}.
\newblock \bibinfo{journal}{\emph{Proceedings of the ACM on Human-Computer
  Interaction}} \bibinfo{volume}{3}, \bibinfo{number}{CSCW}
  (\bibinfo{year}{2019}), \bibinfo{pages}{1--28}.
\newblock
\showISSN{2573-0142}


\bibitem[\protect\citeauthoryear{Lu~Jia, Shen, Shen, Fu, and Peng}{Lu~Jia
  et~al\mbox{.}}{2019}]%
        {lu2019user}
\bibfield{author}{\bibinfo{person}{Adele Lu~Jia}, \bibinfo{person}{Xiaoxue
  Shen}, \bibinfo{person}{Siqi Shen}, \bibinfo{person}{Yongquan Fu}, {and}
  \bibinfo{person}{Liwen Peng}.} \bibinfo{year}{2019}\natexlab{}.
\newblock \showarticletitle{User donations in a user generated video system}.
  In \bibinfo{booktitle}{\emph{Companion Proceedings of The 2019 World Wide Web
  Conference}}. \bibinfo{pages}{1055--1062}.
\newblock


\bibitem[\protect\citeauthoryear{McMillan and Chavis}{McMillan and
  Chavis}{1986}]%
        {mcmillan1986sense}
\bibfield{author}{\bibinfo{person}{David~W McMillan} {and}
  \bibinfo{person}{David~M Chavis}.} \bibinfo{year}{1986}\natexlab{}.
\newblock \showarticletitle{Sense of community: A definition and theory}.
\newblock \bibinfo{journal}{\emph{Journal of community psychology}}
  \bibinfo{volume}{14}, \bibinfo{number}{1} (\bibinfo{year}{1986}),
  \bibinfo{pages}{6--23}.
\newblock


\bibitem[\protect\citeauthoryear{new.qq.com}{new.qq.com}{2021}]%
        {caibao}
\bibfield{author}{\bibinfo{person}{new.qq.com}.}
  \bibinfo{year}{2021}\natexlab{}.
\newblock \bibinfo{booktitle}{\emph{Financial report of Bilibili 2021Q1: The
  revenue was nearly 4 billion, and the monthly active users exceeded 200
  million, but it was difficult to prevent a loss}}.
\newblock
\urldef\tempurl%
\url{https://new.qq.com/omn/20210513/20210513A0CL1E00.html}
\showURL{%
Retrieved July 13, 2021 from \tempurl}


\bibitem[\protect\citeauthoryear{Pintrich}{Pintrich}{2000}]%
        {pintrich2000role}
\bibfield{author}{\bibinfo{person}{Paul~R Pintrich}.}
  \bibinfo{year}{2000}\natexlab{}.
\newblock \showarticletitle{The role of goal orientation in self-regulated
  learning}.
\newblock In \bibinfo{booktitle}{\emph{Handbook of self-regulation}}.
  \bibinfo{publisher}{Elsevier}, \bibinfo{pages}{451--502}.
\newblock


\bibitem[\protect\citeauthoryear{pomofocus.io}{pomofocus.io}{2020}]%
        {pomofocus}
\bibfield{author}{\bibinfo{person}{pomofocus.io}.}
  \bibinfo{year}{2020}\natexlab{}.
\newblock \bibinfo{booktitle}{\emph{Pomofocus}}.
\newblock
\urldef\tempurl%
\url{https://pomofocus.io}
\showURL{%
Retrieved September 12, 2020 from \tempurl}


\bibitem[\protect\citeauthoryear{rescuetime.com}{rescuetime.com}{2020}]%
        {rescuetime}
\bibfield{author}{\bibinfo{person}{rescuetime.com}.}
  \bibinfo{year}{2020}\natexlab{}.
\newblock \bibinfo{booktitle}{\emph{Rescue time}}.
\newblock
\urldef\tempurl%
\url{http://rescuetime}
\showURL{%
Retrieved July 16,2020 from \tempurl}


\bibitem[\protect\citeauthoryear{Rooksby, Asadzadeh, Rost, Morrison, and
  Chalmers}{Rooksby et~al\mbox{.}}{2016}]%
        {rooksby2016personal}
\bibfield{author}{\bibinfo{person}{John Rooksby}, \bibinfo{person}{Parvin
  Asadzadeh}, \bibinfo{person}{Mattias Rost}, \bibinfo{person}{Alistair
  Morrison}, {and} \bibinfo{person}{Matthew Chalmers}.}
  \bibinfo{year}{2016}\natexlab{}.
\newblock \showarticletitle{Personal tracking of screen time on digital
  devices}. In \bibinfo{booktitle}{\emph{Proceedings of the 2016 CHI conference
  on human factors in computing systems}}. \bibinfo{pages}{284--296}.
\newblock


\bibitem[\protect\citeauthoryear{Schunk and Zimmerman}{Schunk and
  Zimmerman}{2012}]%
        {schunk2012motivation}
\bibfield{author}{\bibinfo{person}{Dale~H Schunk} {and}
  \bibinfo{person}{Barry~J Zimmerman}.} \bibinfo{year}{2012}\natexlab{}.
\newblock \bibinfo{booktitle}{\emph{Motivation and self-regulated learning:
  Theory, research, and applications}}.
\newblock \bibinfo{publisher}{Routledge}.
\newblock


\bibitem[\protect\citeauthoryear{Settle and Steinbach}{Settle and
  Steinbach}{2016}]%
        {settle2016improving}
\bibfield{author}{\bibinfo{person}{Amber Settle} {and} \bibinfo{person}{Theresa
  Steinbach}.} \bibinfo{year}{2016}\natexlab{}.
\newblock \showarticletitle{Improving retention and reducing isolation via a
  linked-courses learning community}. In \bibinfo{booktitle}{\emph{Proceedings
  of the 17th Annual Conference on Information Technology Education}}.
  \bibinfo{pages}{34--39}.
\newblock


\bibitem[\protect\citeauthoryear{Southern}{Southern}{2020}]%
        {youtube1}
\bibfield{author}{\bibinfo{person}{Matt Southern}.}
  \bibinfo{year}{2020}\natexlab{}.
\newblock \bibinfo{booktitle}{\emph{Google Lists 5 Key Trends Shaping Consumer
  Behavior Amid COVID-19.}}
\newblock
\urldef\tempurl%
\url{https: //www.searchenginejournal.com/google-lists-5-key-trends-shaping-
  consumer-behavior-amid-covid-19/368874/#close}
\showURL{%
Retrieved July 14, 2021 from \tempurl}


\bibitem[\protect\citeauthoryear{stayfocusd.en.softonic.com}{stayfocusd.en.softonic.com}{2020}]%
        {stayfocused}
\bibfield{author}{\bibinfo{person}{stayfocusd.en.softonic.com}.}
  \bibinfo{year}{2020}\natexlab{}.
\newblock \bibinfo{booktitle}{\emph{StayFocusd}}.
\newblock
\urldef\tempurl%
\url{https: //stayfocusd.en.softonic.com}
\showURL{%
Retrieved July 16, 2020 from \tempurl}


\bibitem[\protect\citeauthoryear{Sun, Wang, and Rosson}{Sun
  et~al\mbox{.}}{2019}]%
        {sun2019distance}
\bibfield{author}{\bibinfo{person}{Na Sun}, \bibinfo{person}{Xiying Wang},
  {and} \bibinfo{person}{Mary~Beth Rosson}.} \bibinfo{year}{2019}\natexlab{}.
\newblock \showarticletitle{How Do Distance Learners Connect?}. In
  \bibinfo{booktitle}{\emph{Proceedings of the 2019 CHI Conference on Human
  Factors in Computing Systems}}. \bibinfo{pages}{1--12}.
\newblock


\bibitem[\protect\citeauthoryear{Tang, Venolia, and Inkpen}{Tang
  et~al\mbox{.}}{[n.d.]}]%
        {44}
\bibfield{author}{\bibinfo{person}{John~C Tang}, \bibinfo{person}{Gina
  Venolia}, {and} \bibinfo{person}{Kori~M Inkpen}.}
  \bibinfo{year}{[n.d.]}\natexlab{}.
\newblock \showarticletitle{Meerkat and periscope: I stream, you stream, apps
  stream for live streams}. In \bibinfo{booktitle}{\emph{Proceedings of the
  2016 CHI Conference on Human Factors in Computing Systems}}.
  \bibinfo{pages}{4770--4780}.
\newblock


\bibitem[\protect\citeauthoryear{Townsend, Kim, and Mesquita}{Townsend
  et~al\mbox{.}}{2014}]%
        {45}
\bibfield{author}{\bibinfo{person}{Sarah~SM Townsend},
  \bibinfo{person}{Heejung~S Kim}, {and} \bibinfo{person}{Batja Mesquita}.}
  \bibinfo{year}{2014}\natexlab{}.
\newblock \showarticletitle{Are you feeling what I’m feeling? Emotional
  similarity buffers stress}.
\newblock \bibinfo{journal}{\emph{Social Psychological and Personality
  Science}} \bibinfo{volume}{5}, \bibinfo{number}{5} (\bibinfo{year}{2014}),
  \bibinfo{pages}{526--533}.
\newblock
\showISSN{1948-5506}


\bibitem[\protect\citeauthoryear{Whittaker, Kalnikaite, Hollis, and
  Guydish}{Whittaker et~al\mbox{.}}{2016}]%
        {whittaker2016don}
\bibfield{author}{\bibinfo{person}{Steve Whittaker}, \bibinfo{person}{Vaiva
  Kalnikaite}, \bibinfo{person}{Victoria Hollis}, {and} \bibinfo{person}{Andrew
  Guydish}.} \bibinfo{year}{2016}\natexlab{}.
\newblock \showarticletitle{'Don't Waste My Time' Use of Time Information
  Improves Focus}. In \bibinfo{booktitle}{\emph{Proceedings of the 2016 CHI
  Conference on Human Factors in Computing Systems}}.
  \bibinfo{pages}{1729--1738}.
\newblock


\bibitem[\protect\citeauthoryear{www.huiian.com}{www.huiian.com}{2021}]%
        {timing}
\bibfield{author}{\bibinfo{person}{www.huiian.com}.}
  \bibinfo{year}{2021}\natexlab{}.
\newblock \bibinfo{booktitle}{\emph{Timing}}.
\newblock
\urldef\tempurl%
\url{http://www.huiian.com/}
\showURL{%
Retrieved July 13, 2021 from \tempurl}


\end{thebibliography}

\end{document}